\documentclass[11pt,twoside]{article}
\usepackage{asp2010}

\resetcounters

\aspvolume{442}
\aspcpryear{2011} 
\aspvoltitle{Astronomical Data Analysis Software and Systems XX} 
\aspvolauthor{Ian N. Evans, Alberto Accomazzi, Douglas J. Mink, and Arnold H. Rots, eds.} 

\begin{document}

\title{Semantic Interlinking of Resources in the Virtual Observatory Era}
\author{Alberto Accomazzi and Rahul Dave}
\affil{Harvard-Smithsonian Center for Astrophysics, 60 Garden Street,
Cambridge, MA 02138, USA}

\begin{abstract}
In the coming era of data-intensive science, it will be increasingly
important to be able to seamlessly move between scientific results, 
the data analyzed in them, and the processes used to produce
them.  As observations, derived data products, publications, and
object metadata are curated by different projects and archived in
different locations, establishing the proper linkages between these
resources and describing their relationships becomes an essential
activity in their curation and preservation.  

In this paper we describe initial efforts to create a semantic
knowledge base allowing easier integration and linking of the body of
heterogeneous astronomical resources which we call the Virtual 
Observatory (VO).  The ultimate goal of this effort is the creation of a
semantic layer over existing resources, allowing applications to
cross boundaries between archives.  The proposed approach follows the
current best practices in Semantic Computing and the architecture of
the web, allowing the use of off-the-shelf technologies and providing
a path for VO resources to become part of the global web of linked data.
\end{abstract}

\section{Introduction}

The explosion of content on the web has been partially tamed by the availability
of services that aim to organize and link resources in ways that 
allow end-users to locate, filter, and rank the available resources.
The enormous success of Google and its pagerank algorithm is mainly
due to its capability of using the architecture of the web to organize
this content, thus demonstrating that
successful web-based information systems need not only 
take into account the content of the resources it knows about,
but also the kinds of connections between them.  

In the commercial world, there are a number of popular 
websites that provide extremely useful services based on 
organizing and presenting information in novel ways which enhance
the discovery process.  Some of the enabling techniques used by
such sites are auto-suggest services, display of ``facets'' to allow
narrowing or broadening of search results, ranking by different
criteria, personalization and recommendations.  
When locating information on the web through one of these
services, the current user expectation is that it not only 
be available through an intuitive interface, 
but also that it be organized in an efficient way, and that relevant
content be only one click away.

These expectations are understandably also present when a scientist
uses web-based services to access resources and data for research
activities.  With the proliferation of scientific digital data 
becoming available from different web-based science archives, it is
essential for information providers to think of their content and 
services as being part of a network of interconnected science products.
As such, their effective discovery and re-use will be enhanced by
portals and search engines that index and expose the context and 
properties of these products through the appropriate interfaces.

Any system supporting resource discovery in astronomy will need to be
built upon our community's distributed environment.
Publications, now
completely in digital format, are published worldwide, but their metadata
is collected and indexed in one single database, the ADS.
Similarly, metadata characterizing Astronomical Objects is
collected by three projects, SIMBAD, Vizier and NED.
While these projects provide a centralized, well-curated 
access to their comprehensive databases of
literature and object metadata, the same is not true for observational datasets.
Observational data
and their basic metadata are stored in a number of archives 
and are usually partitioned based on their observational wavelength
or the observatory which was used to collect them.  
Given the fact that these data are stored in heterogeneous archives and
accessible through interfaces which are very much tied to the underlying data
model, no effective discovery mechanism exists today over this body of data.
While services have been built implementing federated positional searches over 
the contents of data archives, the challenge of providing
a single search paradigm over such an heterogeneous set of data products
has proven difficult to solve in a satisfactory way.

In addition to the problem of ubiquitous discovery and access to datasets,
data preservation principles 
require that we capture, curate, and connect all of the 
activities and digital data products which are part 
of the typical research workflows in astronomy.
In order to support the principle of repeatability of the scientific 
process, it is critical that all artifacts created during a 
scientist's research activity be properly preserved and described
\citep{O05_1_pepe}.  In addition, 
provenance of data used, both between publications and data, and also 
between high-level data products and raw datasets is critical to the reproduction 
of scientific results by others.
Documenting provenance of evidence and conclusions has been done
sporadically and in ad-hoc ways at best, 
but the coming flood of multi-terabyte per night 
data sets require that we adopt best practices and 
frameworks that help us do this efficiently 
and automatically.  

This paper presents work currently being carried out within the US
Virtual Astronomical Observatory (VAO) Data Curation and Preservation 
efforts to create an infrastructure 
supporting curation, discovery and access to VAO resources.  The two main objectives of the
project are to capture and describe the linkages between data and publications
and to capture and describe as much as possible the lifecycle of the research
process, thus enabling us to track the provenance of both data and
publication assets produced by researchers.
Both of these goals contribute to achieving our end goal: creating services
enabling discovery of Virtual Observatory resources via a seamless search
over bibliographic and observational metadata.

\section{Semantics}

In order to provide the proper infrastructure for our project, we
rely on the current best practices and technologies used 
in semantic computing \citep{O05_1_hendler}.  These provide us with 
formal models to uniquely naming resources, concepts, and their
relationships; frameworks to represent and store them in
databases; and standard languages to query and infer over this
knowledge base.
In this section we describe how our project takes advantage 
of these well-established techniques to achieve its goals:
first we identify the resources in our research lifecycle, then we 
model their relationships, and finally we describe them in
a formal way.

\subsection{Resources}

The linkage between astronomical data and publications is complex. 
Data may be used to reach conclusions, and this process is published in papers. 
But data are also measured in order to identify and characterize the celestial
objects which generated the observed signal.  
These Astronomical Objects are then studied by other papers, and additional data 
taken to reach further conclusions about their nature.
Thus, there is a triangle of concepts to consider: Publications, Data, and Objects
(see Figure 1).
Given any instance from one of such concepts, one would like to be able to 
describe (and later discover)
all the possible linkages to the other two, across all known datasets, papers, 
and astronomical objects. 
\begin{figure}[!ht]
\plotone{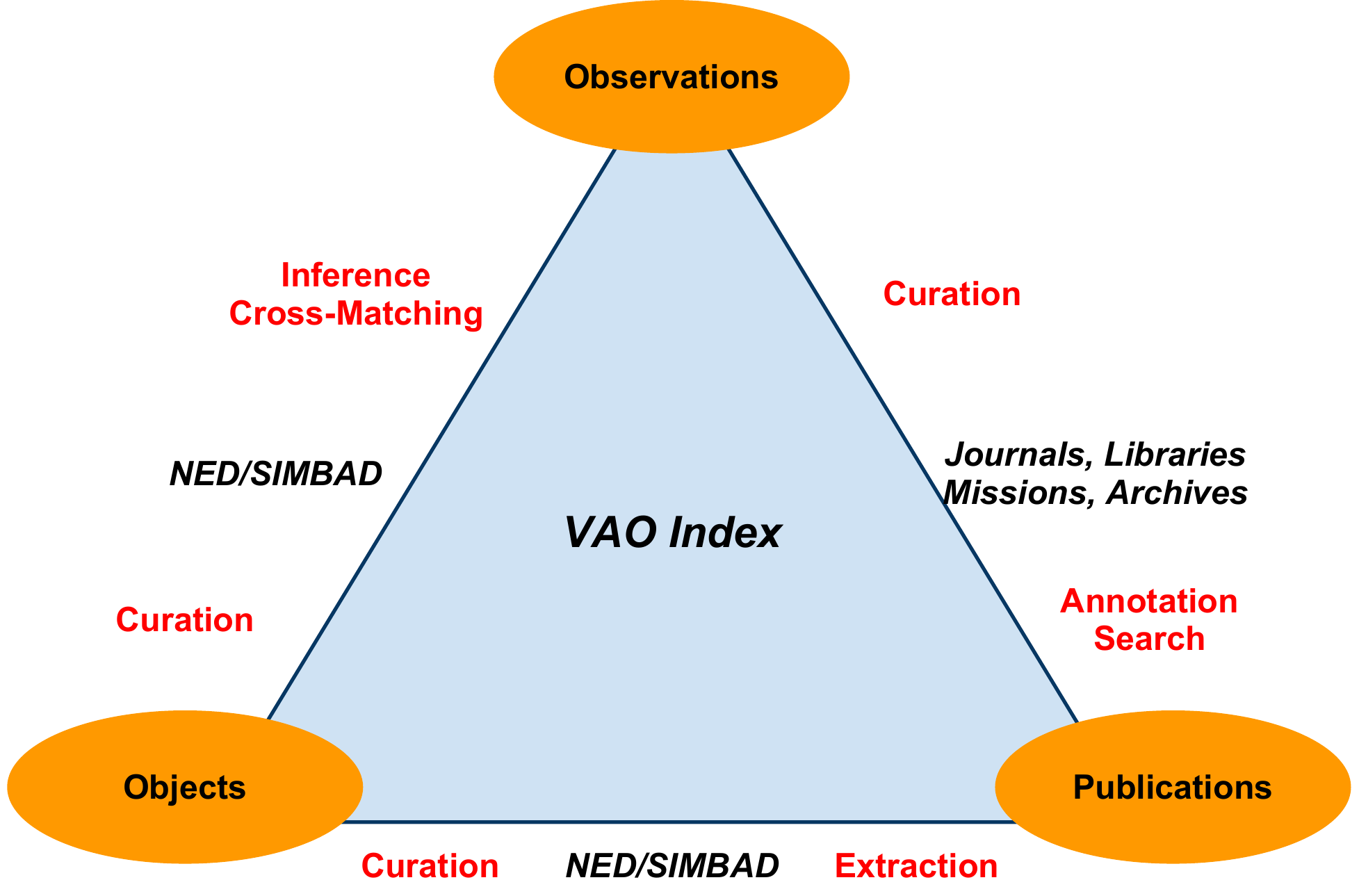}
\caption{Relationships between Publications, Objects, Observations and
the corresponding major actors in the curating process and 
their activities (in red).}
\end{figure}

For example, assume we want to know all papers written about a particular galaxy, 
say M31, and all datasets known about it.  Or, given the Chandra COUP dataset, 
we want to know all known astronomical objects in the footprint of the dataset, 
as well as all papers written using COUP. 
There are further products of these linkages: all datasets sharing overlapping 
footprints, and all papers written about objects in these footprints.

As mentioned earlier, the linkages between publications and Astronomical Objects 
are well curated.
The curation for the linkages between data and objects, 
and between data and publications currently relies on the heroic efforts of 
individual librarians and archivists working at a number of different 
institutes.   Our efforts will leverage
on their work to provide a centralized repository of these links \emph{across}
multiple missions and archives.  
At the same time, we intend to make their work easier by creating an 
infrastructure to simplify the curation process.
Eventually we hope to leverage on other VO efforts and encourage direct participation 
from researchers in identifying linkages between their publications and datasets
described therein.

\subsection{Ontologies}
  
In order to capture the research lifecycle of astronomers, we make use of
formal tools to model the activities and artifacts involved in this process.  
These include: writing a proposal applying to a grant, 
securing funding, making observations, analyzing datasets, creating 
high-level data products, finding and characterizing objects, and writing papers. 
We do so in layers, at each level creating one or more \textbf{Ontologies} 
to describe the concepts and activities within the layer.
We first start with the fundamentals of the 
Scientific process, creating an ontology called \textbf{VAOBase}. 
We build on that an observational ontology, \textbf{VAOObsv}, which describes 
observations and their associated datasets. We also build an ontology for 
publications, \textbf{VAOBib}, which relates to the other two ontologies.

An ontology is a formal representation of the concepts within a domain of 
knowledge \citep{O05_1_hendler}, and of the relationships between these concepts. 
For example, an
\emph{Observation} is a subtype of a \emph{ScienceProcess} \textbf{Class}, 
and it results in a \emph{DataProduct} \textbf{Class}. 
We represent this linkage as a property named \emph{hasDataProduct}. 
We then say that an instance of the \emph{Observation} class \emph{hasDataProduct} an 
instance of the \emph{DataProduct} class.
We define ontologies in a formal language known as OWL (Ontology Web Language,
\citet{O05_1_mcguinness}), 
which is itself defined in a simpler formal language 
called RDF (Resource Description Framework\footnote{\url{http://www.w3.org/RDF/}}).
RDF is widely used on the web, and its use has led to 
the development of a parallel web of resources that can be linked to each other, 
and whose descriptions are machine readable, 
called the Semantic Web\footnote{\url{http://en.wikipedia.org/wiki/Semantic_Web}}.
Since RDF provides typed links between resources, every site that publishes RDF 
contributes to a large, world-wide graph over which computations can be performed. 
Such computations include relational database like queries on the graph using 
an analog of SQL called SPARQL, as well as the inferring of relationships 
between resources from existing relationships in the graph.

We have chosen to use industry standard RDF and OWL technologies since 
these are widely deployed, and have very good tool support.  
Furthermore, we can make use of a number of existing excellent 
ontologies to build upon.  
These include the Provenance, Authoring, and Versioning ontology from the 
SWAN Project\footnote{\url{http://swan.mindinformatics.org/ontology.html}}
which provides a basis for all provenance related activity 
in our ontologies.
We also use the FABio and CiTO ontologies from the 
Semantic Publishing and Referencing 
Ontologies\footnote{\url{http://opencitations.wordpress.com/}}
which provide a way for typing the different kinds of publications and 
citations respectively.  
For Astronomy semantics, we utilize the IVOA SKOS 
vocabularies for astronomical 
keywords\footnote{\url{http://www.ivoa.net/Documents/latest/Vocabularies.html}}, 
as well as the CDS vocabulary 
for Astronomical Objects and their variability 
types \citep{O05_1_derriere}.

Our model of Observations and Data Products follows that of 
the Common Archive Observation Model (CAOM, \cite{O05_1_caom}). 
Wherever possible, we have chosen to track existing, deployed standards.
Datums, datasets, and their associated observations are described by 
metadata properties such as position, URI, flux data, band, Instruments used, etc. 
We have chosen the metadata properties we wish to model and their names 
following the ObsCore specification from the ObsTAP 
project\footnote{\url{http://www.ivoa.net/cgi-bin/twiki/bin/view/IVOA/ObsDMCoreComponents}}. 
ObsCore is rapidly gaining steam amongst archives as a minimal, 
simple standard to provide 
ADQL \footnote{\url{http://www.ivoa.net/Documents/latest/ADQL.htm}}
compatible querying of data product 
metadata, and we intend to ride on its coattails.

\subsection{Representing the Research Lifecycle in Astronomy}

In this section we illustrate, by way of example, how the formal
tools described above can be used to represent scientific assets, 
their relationships, and research activities performed on them.
A schematic representation
of the main concepts and relationships can be found in Figure 2, and
a narrative of some of these activities is given below.
\begin{figure}[!ht]
\plotone{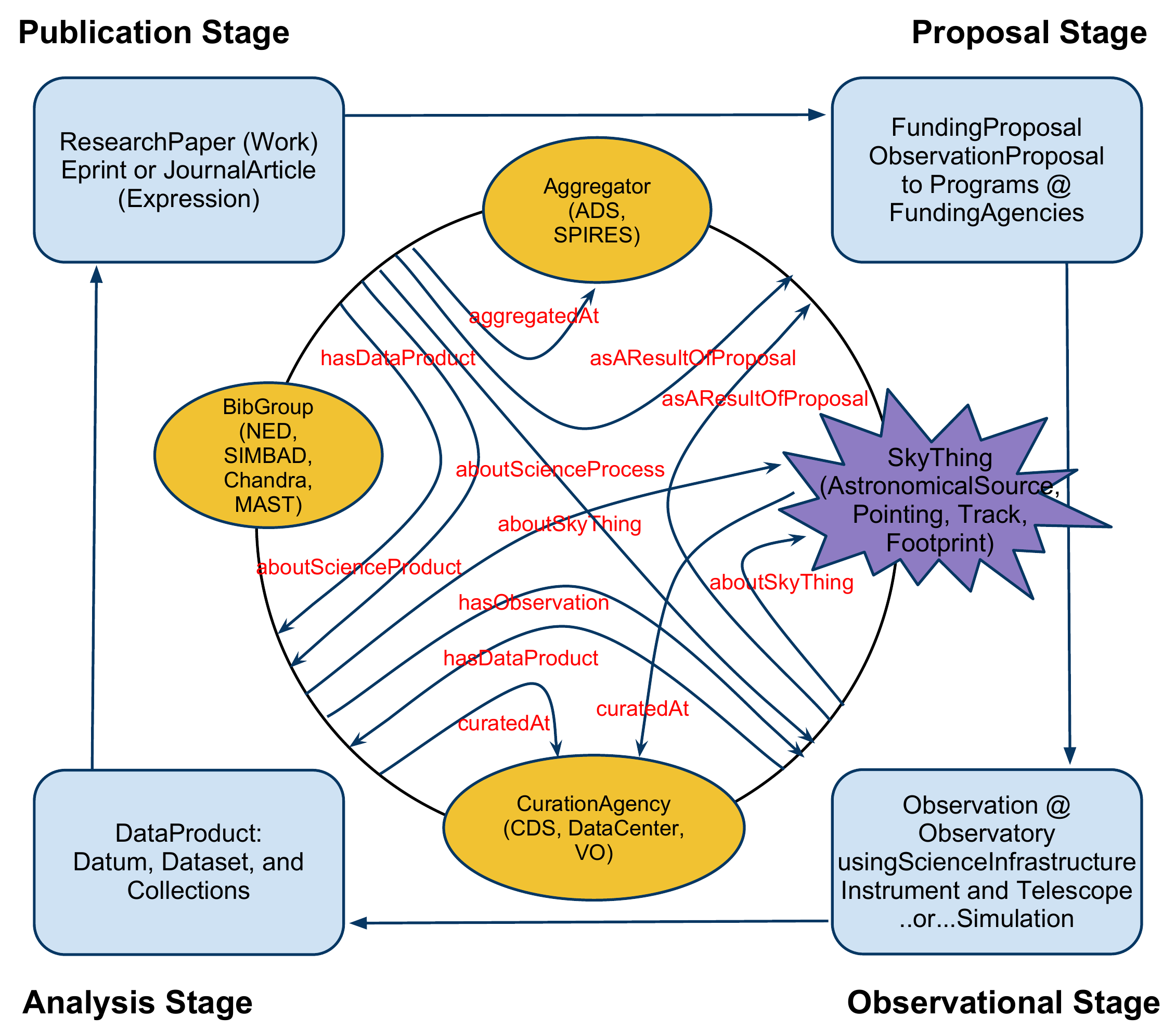}
\caption{A model of the Research Lifecycle in Astronomy, 
showing some of the classes in our three ontologies, and some of the links 
between instances of these classes (created as ObjectProperties in OWL). 
For example, an instance of an Observation may (or may not) have the property 
\emph{asAResultOfProposal} whose range is an instance of the class 
\emph{ObservationProposal}.}
\end{figure}

We submit \emph{Proposal}s for funding and \emph{ObservationProposal}s 
for time to \emph{Program}s and \emph{ObservationProgram}s at 
\emph{FundingAgency}s and \emph{ScienceInfrastructureAgency}s respectively. 
Upon grant of the proposals we carry out a type of \emph{ScienceProcess} 
called \emph{Observation} at \emph{ScienceInfrastructure} such as 
\emph{Observatory}s using \emph{Instrument}s and \emph{Telescope}s. 
We then carry out \emph{Analysis} of the observations  leading to the 
production of \emph{DataProduct}s. 
Further analysis and possibly \emph{Simulation}s, also examples of 
science processes, are carried out leading to the production of 
\emph{WrittenProduct}s such as reports or papers.

Observations taken on the sky may be known \emph{AstronomicalSource}s 
at known \emph{Position}s, as identified by one or more 
\emph{CurationAgency}s such as the CDS, or of a random \emph{Pointing}, 
\emph{Track}, or \emph{FootPrint} on the sky.  
Observations may be \emph{SimpleObservation}s, which correspond to photons 
collected in one time interval or \emph{ComplexObservation}s such as 
multi-point skews, grid observations, etc. A piece of data from a simple 
observation is called a \emph{Datum}, e.g., the FITS file corresponding 
to a single exposure. 
Multiple simple observations (more precisely their data) may be combined 
into a \emph{Dataset}, such as a mosaic, or light curve. 
\emph{ComplexObservation}s too are represented by datasets. 
Both datum and dataset are types of \emph{SingularDataset}s, 
which might be combined together to create \emph{CompositeDataset}s such as 
cartouches of all files associated with a given astronomical source.

Publications are described in our ontologies using FABIO's support for FRBR
(Functional Requirements for Bibliographic 
Records\footnote{\url{http://archive.ifla.org/VII/s13/frbr/frbr_current_toc.htm}}). 
FRBR advocates tracking \emph{Work} through its various \emph{Expression}s, 
and the \emph{Manifestation}s of these expressions. 
For example, the work in case may be a \emph{ResearchPaper} on spiral galaxies. 
This paper is expressed as a \emph{JournalArticle}, and before this article 
is ever published, as an \emph{Eprint} on the Arxiv 
site\footnote{\url{http://arxiv.org}}.
Manifestations of this paper are the various formats in which the article 
is available, at various online \emph{Aggregator}s, or in printed form. 
Such a research paper represents a \emph{WrittenProduct} about the data products, 
observations and analysis. 

The links from Publications to Data, and from Publications and Data to 
Proposals are maintained by \emph{BibGroup}s at various 
institutions.  These links are captured in our ontologies by 
properties such as \emph{aboutScienceProcess}, \emph{aboutScienceProduct}, 
\emph{underProgram}, \emph{asAResultOfProposal} and \emph{hasDataProduct} 
whose domain is the Work or Expression, or even the 
\emph{Observation} or \emph{AstronomicalSource} at hand.  
These linkages constitute the key part of our project.

It is probably obvious by now than any such database of such resources 
and linkages is incomplete.  Here the usage of semantic technology 
compared to relational technology shines:  we only need to assert the 
properties we know about. 
Nevertheless, our framework has been designed so that all the crucial 
entities and their properties can be captured according to the model
at any point in time.  As an example, we intend to use text mining techniques
at a later date to search the fulltext literature for grant numbers, 
program names, and organizations.
There are many other terms defined in our ontologies and the ontologies that 
they depend upon. These can be examined in more detail in our code 
repository\footnote{\url{https://github.com/rahuldave/ontoads}}.

\section{Infrastructure and Applications}

In the previous section we discussed the concepts that our ontologies 
capture, and the languages we represent these concepts in, RDF and OWL. 
The reason for using these languages is the vast 
infrastructure available as open source software for the semantic web.
The purpose of our server and database infrastructure is to:
 provide a linked data endpoint to various astronomical resources 
and the relationships between them;
enable the querying and inferencing on this graph of resources and relationships;
index certain key resources and relationships in order to provide 
a fast query interface over selected properties of publications, datasets, 
and astronomical objects;
enable application such as search and discovery engines, and faceted browsers
of astronomical resources to be built, so as to deliver services to end users;
enable applications to be built which will help future identification of 
data-publication linkages, and provide these services to bibliographic groups 
at different astronomy institutions, as well as directly to astronomers.

We intend to create an indexed database 
of publications, datasets, and their relationships
to provide an effective infrastructure for resource discovery, leveraging
on ADS's expertise in metadata and full-text indexing.
Bibliographic metadata will be incorporated into the knowledge base from
the ADS database.
Integration of object metadata and linkages will be accomplished 
utilizing the astronomical object databases maintained by NED and SIMBAD.
Observational metadata will be incorporated from a number of collaborators
at the CDS, Chandra, NED and MAST who maintain
curated connections between datasets and publications.

\subsection{Server Infrastructure}

To store RDF statements, we use a database system called a triplestore,
and have selected the open source Sesame\footnote{\url{http://www.openrdf.org/index.jsp}} 
as the DBMS.
The triplestore stores statements, creates indexes on 
some subjects, objects, and predicates, and provides SPARQL and 
RESTian\footnote{\url{http://en.wikipedia.org/wiki/Representational_State_Transfer}}
interfaces to resources and simple queries. 
Additionally, Sesame stores triples with a context, which may then be used 
to track transactional additions and removals from the database.

However, because a triplestore has no knowledge of the structure of relationships 
in the data, it provides slow performance in the common search cases, such as
finding the datasets associated with a publication, for example.
To provide fast results which can then be faceted, we use 
SOLR\footnote{\url{http://lucene.apache.org/solr/}}
as an indexing server in front of the 
triplestore.
This allows us to have a two-tier system, where complex SPARQL or 
subject/object/predicate queries are handed over to 
Sesame, while SOLR serves the more common search cases with real fast indices. 
Furthermore, since SOLR provides faceting out of the box, we can write 
user interfaces for our application, once we index the
properties we wish to filter upon.

Finally, a web service written in python is used to make choices as to 
which server to query and proxy, manage authentication, run federated searches
to SIMBAD and NED, and handle any additional 
features that a user-facing application requires.
The triplestore is currently accessible via the SESAME API and SPARQL query language, 
using our python library code. 
We use it in our data pipeline to populate the SOLR server, 
and to inferentially add data into it.
We are planning to use this infrastructure in the core pipeline
for ingesting publications from the ADS to normalize
author and organization names, to keep track of the linkages between 
papers and proposals, and to track the provenance of publications.

The triplestore has been populated with a select subset of bibliographic data 
from ADS and makes use of an object cache automatically populated as SIMBAD and NED
are queried.
For observational data, our strategy is to ingest metadata from 
larger archives (starting from Chandra and MAST), and make our way to the smaller ones. 
Chandra data is complex and we have collaborated with the Chandra Archive team 
to convert their metadata into RDF.
Because our observation model is based on ObsCore and CAOM, we will
be able to ingest data from any mission which publishes metadata in ObsCore 
compatible tables.  This is how we will be tackling most of the data from MAST.

\subsection{Applications}

A first prototype user interface is being developed in javascript with jquery, AJAXSolr
and our own custom code which talks to the backend SOLR indexing server, Sesame triplestore, 
and the python web service. 
This user interface makes use of AJAX to pull metadata from the server in the background 
while the interface is being manipulated.
\begin{figure}[!ht]
\fbox{\plotone{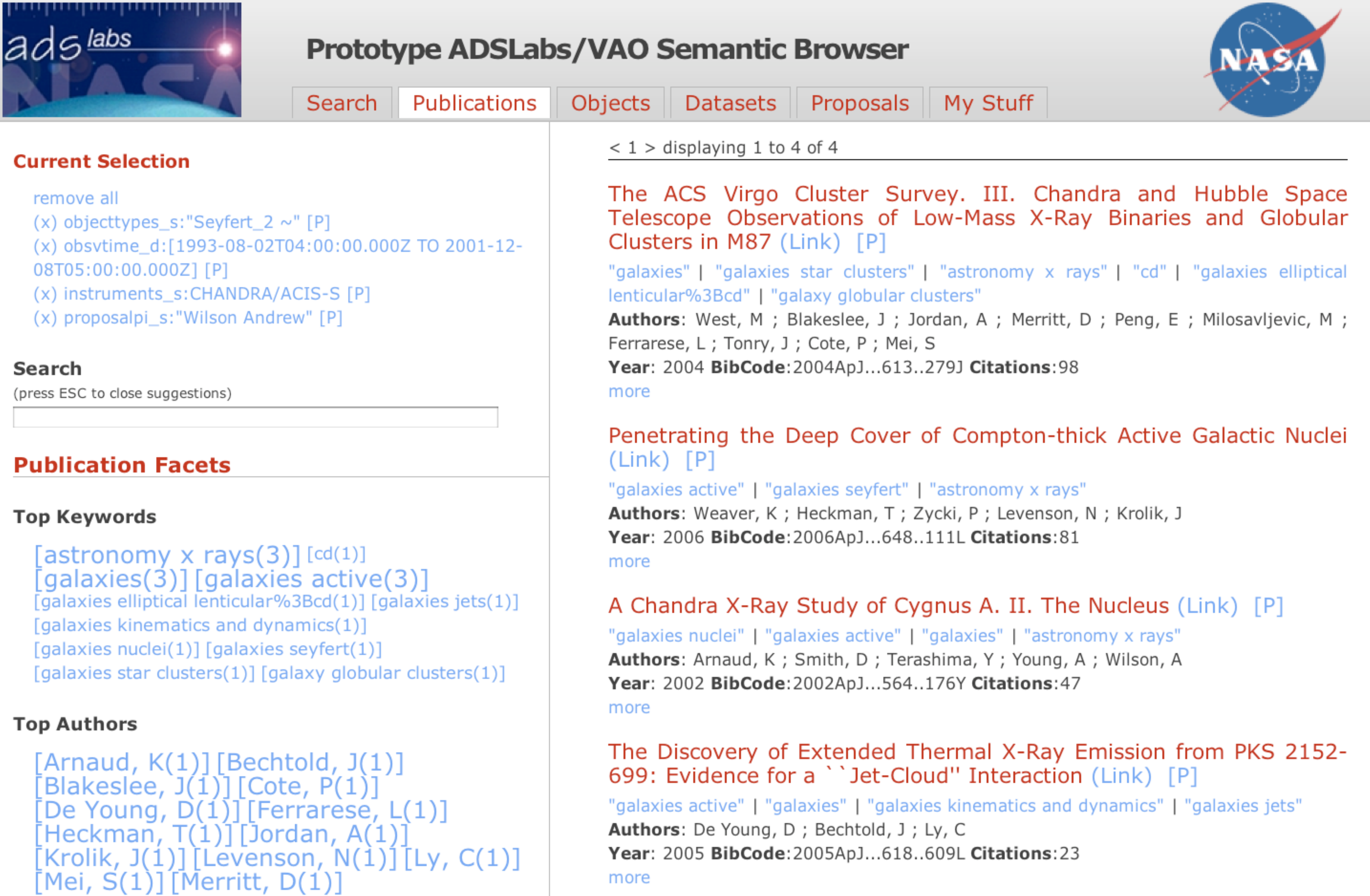}}
\caption{A prototype of a faceted search on publications,
with filtering via observational and object metadata}
\end{figure}

In the screenshot depicted in 
Figure 3, publications are being faceted by various metadata belonging to the 
datasets used in them, the objects described within, and the proposals used to 
fund the research and make observations.
Clicking on any facet link will filter the publication set by that facet in addition to 
the facets already chosen; clicking a ‘P’ (or pivot) link will change to a view in 
which the publications are filtered by that facet only. 
In the figure, we have faceted by Seyfert Objects, observation time, the CHANDRA ACIS-S 
instrument, and selected a particular proposal PI (Andrew Wilson).  
These simple filtering activities lead us to find papers associated with Seyfert research 
proposed by Andrew Wilson in a particular timeframe and with a particular instrument. 
Interestingly, only one of the papers that result from this selection is co-authored by him, 
indicating that these observations have had impact beyond the original intent of the
proposal, a result that would have been difficult to conclude without the support
of this knowledge base.

This interface is being extended to facet datasets, objects, and proposals
in order to provide a generic search and bookmarking capability over all
these resources.   It will be made available as part of the VAO toolset,
the ADS ``Labs'' experimental search interface, and possibly integrated in
the upcoming VAO portal.

\section{Conclusions and Future work}

Our backend server infrastructure and javascript prototype experiments 
accomplish a first goal:
exposing the linkages between objects, datasets, and publications in a natural way, 
thus making it easier for astronomers to explore the space of astronomical 
concepts and phenomena
using an iterative process through an interface which exposes key relationships among them.
The knowledge base and infrastructure we are building is meant to provide support for a 
variety of applications,
some of which we will develop ourselves, with others being contributed by collaborators.
A list of potentially useful applications that we have envisioned include:
\begin{itemize} 
\item {\bf The APOD Browser}: A 3-pane search and exploration browser which will allow 
users to simultaneously browse Astronomical Publications, Objects, and Datasets (APOD). 
The contents of each pane view will change depending on selections in the other panes. 
It will also be possible to pivot on any asset in any pane and see what resources are
available for the other two.  Any search will be bookmarkable and will act as a live search, 
so that additions to our and other mission and archival databases will be immediately 
reflected in the search through a process of notification. 
Thus APOD will serve as a research portfolio tool for graduate students and seasoned 
astronomers alike.  By linking APOD into the VAO portal, we will be able to provide 
one-stop service to users of the VAO.

\item {\bf Annotation Server}: The working of our tool depends largely on the mostly unsung 
efforts of bibliographic groups maintained by multiple archives such as Chandra, MAST, ESO, 
NED, CDS, and ADS.  By combining our triple store with semantic annotation technology 
and the ADS literature full text search, we are in the position to provide infrastructure 
to the maintainers of bibliographic groups to carry on their annotation of literature-data 
and object-literature connections in a more efficient manner, simplifying their curation efforts.

\item {\bf Metrics tool}: By leveraging the efforts of bibliographic groups across multiple 
missions, and by full-text mining of publications, we are also capable of providing a 
queryable infrastructure that links publications to proposals and observations.  
This allows the computation of metrics on the efficacy of observing and funding programs, 
as well as the output of researchers.  This is invaluable information for both funding 
agencies and mission directorates.  Thus user interfaces can be developed which make 
such metric extraction as easy as the faceted browsing of astronomical concepts.

\item {\bf Paper of the Future}: Leveraging on the database of the connections 
from any given publication to the objects studied therein, the datasets used, 
and the proposals that went into the production of the paper, we will be able to 
provide a more wholistic view of the paper, with direct linking to (and in some case 
inline depiction of) datasets, catalogs, objects, SEDs, etc. 
In conjunction with full text searching, the extraction of table, figure, and 
equation assets from the paper, and added encouragement to users to provide enhanced 
publication-data linking themselves, we will be able to provide a very rich view of 
the paper itself.  In addition, we will be able to provide to the users links to 
relevant resources and recommendations based on a variety of criteria, 
such as citations, usage of
data products, objects studied, etc.
\end{itemize}

We have emphasized earlier the dawn of a new age of data-intensive astronomy,
which will require a paradigm shift in the way research is conducted in our
discipline.
The work we have presented in this paper 
is part of the effort to automate and make easier the characterization 
and indexing of scientific resources and their relationships.
Additionally, by capturing and formally describing the linkages from
published research to data used, we will make progress towards the
creation of a digital environment enabling the repeatability
of the scientific process.

\acknowledgements We are grateful to a number of individuals and groups who have 
provided us with the metadata currently indexed in our knowledge base, in
particular Sherry Winkelman (Chandra), Karen Levay (MAST), and the
SIMBAD, NED and ADS teams.  Sherry and members of the VAO collaboration, in particular 
Doug Burke, Matthew Graham and Brian Thomas offered suggestions on a number of topics
related to the development of our Ontologies and technical infrastructure.
We thank Alyssa Goodman and Michael Kurtz for inspiring us to pursue this effort.
This work was supported by the Astrophysics Data
System project which is funded by NASA grant NNX09AB39G, 
Microsoft Research WorldWideTelescope, 
and the Virtual Astronomical Observatory, funded under NSF and NASA grants.

\bibliography{O05_1}

\end{document}